\documentclass[a4,letter]{aa}

\usepackage{graphicx}
\usepackage{txfonts}
\usepackage{multirow}
\usepackage{makecell}
\usepackage{subcaption}
\usepackage{placeins}
\usepackage{array}
\usepackage{booktabs}
\usepackage{xcolor}
\usepackage{helvet}  
\usepackage{ulem}
\usepackage[colorlinks=true,allcolors=blue]{hyperref}

\newcommand{\struct}[1]{}

\begin{document}

\title{Thermal Electrons in an Ultra-Relativistic Shock Shape the Optical Afterglow of GRB 250702F}

\author{
Martin~Jel\'{i}nek\inst{1}\thanks{\email{martin.jelinek@asu.cas.cz}}
\and Annarita~Ierardi\inst{2,3}\thanks{\email{annarita.ierardi@gssi.it}}
\and Filip~Novotn\'{y}\inst{1,4}\thanks{\email{filip.novotny@uni-potsdam.de}}
\and Gor Oganesyan\inst{2,3}\thanks{\email{gor.oganesyan@gssi.it}}
\and Biswajit~Banerjee\inst{2,3}
\and Dimitrios~Giannios\inst{5}
\and Sergey~Karpov\inst{6}
\and Martin~Topinka\inst{7,8}
\and Elias~Kammoun\inst{9}
\and Jan~\v{S}trobl\inst{1}
\and Alberto J. Castro-Tirado\inst{10}
}

\institute{
Astronomical Institute of the Czech Academy of Sciences (ASU-CAS), Fri\v{c}ova 298, 251~65 Ond\v{r}ejov, Czech Republic
\and Gran Sasso Science Institute, Viale F. Crispi 7, I-67100 L'Aquila (AQ), Italy
\and INFN -- Laboratori Nazionali del Gran Sasso, I-67100 L'Aquila (AQ), Italy
\and Department of Theoretical Physics and Astrophysics, Faculty of Science, Masaryk University, Kotl\'{a}\v{r}sk\'{a} 2, Brno, 611~37, CZ
\and Department of Physics and Astronomy, Purdue University, West Lafayette, IN, USA
\and Institute of Physics of the Czech Academy of Sciences, Prague, CZ
\and INAF -- Osservatorio Astronomico di Cagliari, Via della Scienza 5, 09047 Selargius (CA), Italy
\and Charles University, Faculty of Mathematics and Physics, Astronomical Institute, V~Hole\v{s}ovi\v{c}k\'{a}ch 2, Prague, 180~00, CZ
\and Cahill Center for Astronomy \& Astrophysics, California Institute of Technology, 1216 East California Boulevard, Pasadena, CA 91125, USA
\and Instituto de Astrofísica de Andalucía (IAA-CSIC), Glorieta de la Astronomía s/n, Granada, 18008 Spain
}

\date{Received xxx; accepted xxx}

\abstract{
Observing early optical emission from gamma-ray bursts (GRBs) contemporaneous with the MeV prompt emission phase remains rare, requiring rapid-response robotic facilities.
The Ond\v{r}ejov D50 telescope detected the optical counterpart of GRB~250702F at $z=1.520$ only 27.8\,s after trigger, enabling high-cadence monitoring during the brightest prompt emission pulses.
The optical light curve reveals two distinct flares. The first ($30$--$100$\,s) is spectrally consistent with the MeV prompt emission. The second flare ($100$--$1400$\,s) exhibits an unusual morphology ($F_{\nu} \propto t^{-\alpha}$): a rapid rise to a plateau, followed by a steep decay ($\alpha \sim 1.6$) before transitioning to a standard power-law afterglow ($\alpha = 0.79$).
This steep decay phase cannot be explained by nonthermal electrons accelerated at the forward shock, and reverse-shock scenario is disfavored due to the long duration of the flare and the temporal offset from the underlying deceleration time.
We interpret the steep decay as the synchrotron frequency of a thermal (Maxwellian) electron population sweeping through the optical band. Modeling yields a non-thermal energy fraction $\delta \approx 0.8$ with the remaining energy heating electrons at characteristic Lorentz factor $\gamma_{\rm th} \sim$ 900. 
These observations provide evidence for thermal electron signatures in GRB afterglows, consistent with predictions from particle-in-cell simulations of ultra-relativistic collisionless shocks.
}

\keywords{
gamma-ray burst: individual: GRB~250702F --
gamma-ray burst: general --
acceleration of particles --
shock waves --
radiation mechanisms: non-thermal --
techniques: photometric
}

\maketitle
\nolinenumbers

\section{Introduction}

\struct{What are GRBs and afterglows -- for non-GRB audience}
Gamma-ray bursts (GRBs) are among the most luminous transient events in the Universe, produced by ultra-relativistic jets launched after the collapse of massive stars or the merger of compact objects. The prompt emission, lasting from milliseconds to hundreds of seconds, is dominated by MeV gamma-rays originating from internal dissipation within the jet. As the jet decelerates against the circumburst medium, it drives an external shock that produces broadband afterglow emission from radio to TeV $\gamma$-rays, observable for days to months after the burst \citep{MeszarosRees1997,Sari1998}.

\struct{Early optical observations are rare -- state of the art}
Optical observations during or immediately after the prompt emission phase remain rare, as they require rapid-response robotic telescopes capable of repointing within seconds of a satellite trigger. Since the first detection of prompt optical emission in GRB~990123 \citep{Akerlof1999}, such observations have revealed diverse phenomenology. Some bursts show optical variability correlated with MeV pulses, suggesting a common origin in internal dissipation \citep{Vestrand2005,Racusin2008,Beskin2010}. Others exhibit rapidly decaying optical flashes, traditionally interpreted as emission from the reverse shock propagating back into the ejecta \citep{Akerlof1999,Fox2003,Gomboc2008}. High-cadence coverage of the prompt-to-afterglow transition is valuable, as it probes the physics of relativistic shocks during the deceleration phase.

\section{Observations}
\label{sec:data}

GRB~250702F is a long-duration gamma-ray burst at redshift $z = 1.520$ with isotropic-equivalent energy $E_{\rm iso} = (9.9 \pm 1.5) \times 10^{52}$\,erg, detected by \textit{Fermi}/GBM and \textit{Swift}/BAT \citep{GCN40892,GCN40894,GCN40948}. The Ond\v{r}ejov D50 robotic telescope \citep{d50telescope} began observations 27.8\,s after trigger, providing high-cadence optical coverage through the prompt-to-afterglow transition (Appendix~\ref{app:photometry}).

\begin{figure}[t]
    \centering
    \includegraphics{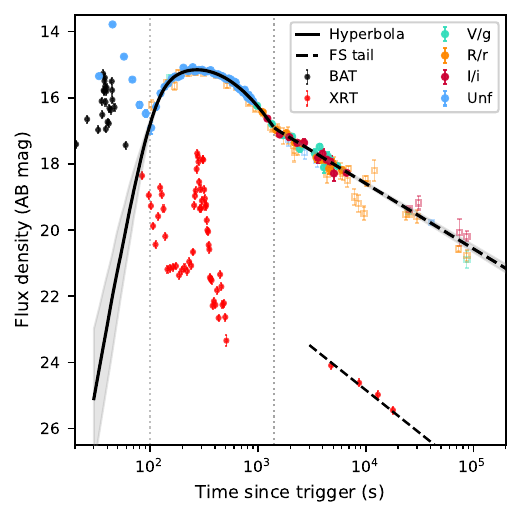}
    \caption{Multi-wavelength light curve of GRB~250702F. Optical data from D50 are shown together with \textit{Swift}/BAT and XRT observations. Multi-filter optical points are color-corrected to the $r$-band using the measured spectral slope. Open symbols represent GCN points. The solid line shows an empirical fit to the optical data (see Section~\ref{sec:lcmodel}).}
    \label{fig:lc}
\end{figure}

Figure~\ref{fig:lc} shows the multi-wavelength light curve. The optical data reveal a two-flare structure: flare~A (30--100\,s) coincides with the brightest prompt emission pulses detected by \textit{Swift}/BAT, while flare~B (100--1400\,s) shows an unusual rise--plateau--steep-decay morphology before transitioning to a standard afterglow at $t \gtrsim 1400$\,s. X-ray flaring detected by \textit{Swift}/XRT during 100--500\,s is notably decoupled from the optical evolution.

\section{Prompt emission}
\label{sec:sed}

\struct{GBM spectral analysis of SP1, SP2}
To probe the origin of optical flare~A (30--100\,s), we performed time-resolved spectral analysis of the contemporaneous \textit{Fermi}/GBM emission (Appendix~\ref{app:gbm}) in two intervals: SP1 (28.5--38.5\,s) and SP2 (39.7--49.7\,s). In both cases, the spectra are well fit by a cutoff power-law model. For SP1, we find a low-energy photon index $\alpha_X = -1.14_{-0.06}^{+0.07}$ and peak energy $E_{\rm p} = 763_{-229}^{+153}$\,keV, while SP2 yields $\alpha_X = -1.49 \pm 0.07$ and $E_{\rm p} > 2$\,MeV.

\struct{Key result: optical consistent with prompt extrapolation}
Extrapolation of the best-fit models to optical frequencies predicts fluxes consistent with the observed D50 measurements (Fig.~\ref{fig:spectra}, bottom left panel). Combined with the temporal coincidence of flare~A with the brightest MeV pulses, this supports a common origin in internal dissipation within the relativistic jet. The optical emission during this phase is not a separate afterglow component but rather the low-energy extension of the prompt spectrum.

\struct{X-ray flares: continued internal activity, optical decoupled}
After $t \sim 100$\,s, the GRB continues flaring in X-rays (Appendix~\ref{app:xrt}), with flux enhancements of $\sim$10--100$\times$ above any underlying power-law decay; the optical emission has by then decoupled and shows no corresponding re-brightenings. Joint XRT--BAT spectral analysis for six time intervals (SP3--SP8), with priors that the extrapolated spectrum not exceed the observed optical flux, yields smoothly broken power-law fits with peak energies $\sim$0.1\,keV to $\sim$4\,keV and photon indices $\alpha_X \approx -0.8$ and $\beta_X \approx -2.2$ to $-2.4$ (Fig.~\ref{fig:spectra}, bottom right panel). These spectral shapes match typical photon indices of MeV prompt emission pulses \citep{Nava2012}, supporting X-ray flares being softer analogs of prompt emission pulses. \textit{Fermi}/LAT upper limits (0.1--1\,GeV) lie above the extrapolated spectra and do not further constrain the spectral shape.

\section{Optical light curve analysis}
\label{sec:lcmodel}

\subsection{Empirical characterization}

\struct{Describe the morphology: rise, plateau, steep decay, standard afterglow}
Beyond $t \sim 100$\,s, the optical light curve shows a rapid rise transitioning to a flat maximum around $t \sim 200$\,s. This is followed by gradual steepening into a decay phase reaching $\alpha \sim 1.6$--$2.0$ (steepest just before $t \sim 1400$\,s), then a sharp break to a shallower power-law decline $\alpha \sim 0.8$ that persists to late times.

\struct{Empirical fit parameters}
We fitted the light curve with an empirical model consisting of a double hyperbola (capturing rise, plateau, and steep decay) joined to a late-time power law constrained by forward-shock closure relations (Appendix~\ref{app:lcfits}). The best-fit parameters are: rise index $\alpha_1 = -3.1^{+0.9}_{-0.6}$, plateau index $\alpha_2 = -0.17^{+0.18}_{-0.29}$, steep decay index $\alpha_3 = 1.8^{+0.8}_{-0.6}$, with transition times at $t_{v,1} = 114^{+14}_{-9}$\,s and $t_{v,2} = 870^{+470}_{-300}$\,s. The break to standard afterglow behavior occurs at $t_b = 1405^{+67}_{-66}$\,s. The late-time decay is well described by electron distribution index of $p = 2.05 \pm 0.04$, yielding $\alpha_4 = 0.79 \pm 0.03$ and spectral slope $\beta = 0.52 \pm 0.02$ via ISM slow-cooling closure relations.

The steep-to-normal decline ($>$800 s) is incompatible with the standard forward shock scenario \citep{Sari1998}. The rapid rise and steep decay are broadly consistent with emission from electrons in the GRB ejecta heated by a reverse shock (e.g., \citep{NakarPiran2004}). However, the presence of a shallow segment is difficult to reconcile with the standard picture of a single, homogeneous shocked shell. Moreover, in the reverse plus external forward shock scenario, the deceleration time of both components are expected to coincide, which is inconsistent with the data (see Appendices~\ref{app:twocomp} and ~\ref{app:archival}).

\struct{Key observation: two transitions, jet break, Gamma0}

\begin{figure*}[t]
    \centering
    \sidecaption
    \includegraphics[width=12cm]{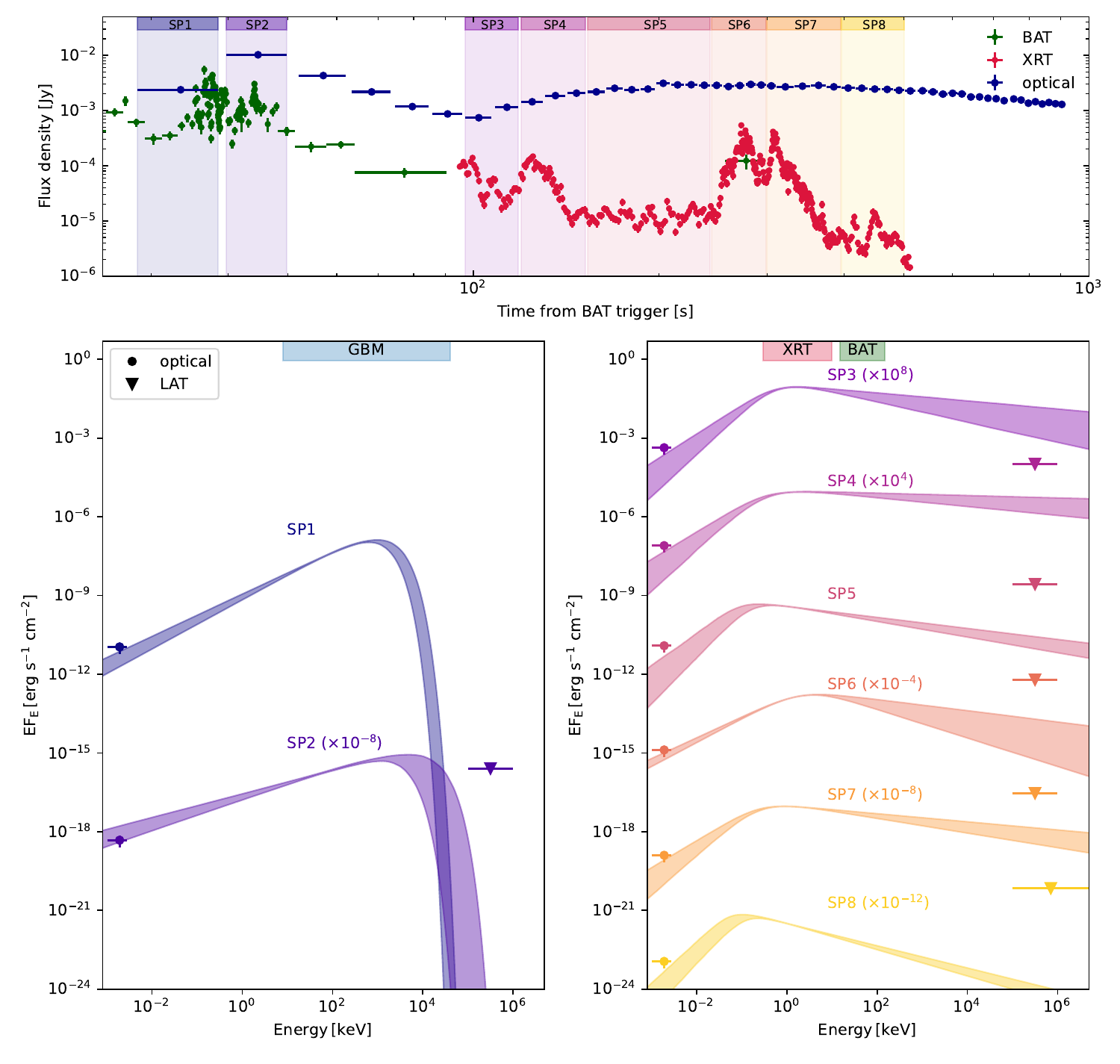}
    \caption{Multi-wavelength light curves (top panel), spectral energy distributions during the prompt phase (bottom left panel) and during the X-ray flares (bottom right panel). The optical data points from D50 and the LAT upper limits are shown together with the best-fit spectral models. During the prompt phase (SP1, SP2), the extrapolated gamma-ray spectrum is consistent with the observed optical flux. During the X-ray flares (SP3--SP8), the optical flux was set as an upper limit for the joint fit.}
    \label{fig:spectra}
\end{figure*}

\struct{Standard interpretations excluded}

\subsection{Thermal electron interpretation}
\label{sec:thermal}

\struct{Hybrid electron model (GS2009) and physical picture}
We consider a forward-shock model with a hybrid electron distribution: a thermal (Maxwellian) component smoothly connected to a non-thermal power-law tail \citep{GS2009}. The thermal population produces a synchrotron peak at frequency $\nu_{\rm th}$, initially above the optical band. As the shock decelerates, $\nu_{\rm th} \propto t^{-3/2}$ sweeps through the optical band, producing the observed plateau followed by steep decay. Once the thermal contribution fades, the standard non-thermal afterglow emerges at $t \sim 1400$\,s.

\begin{figure}[t!]
    \centering
    \includegraphics[width=\columnwidth]{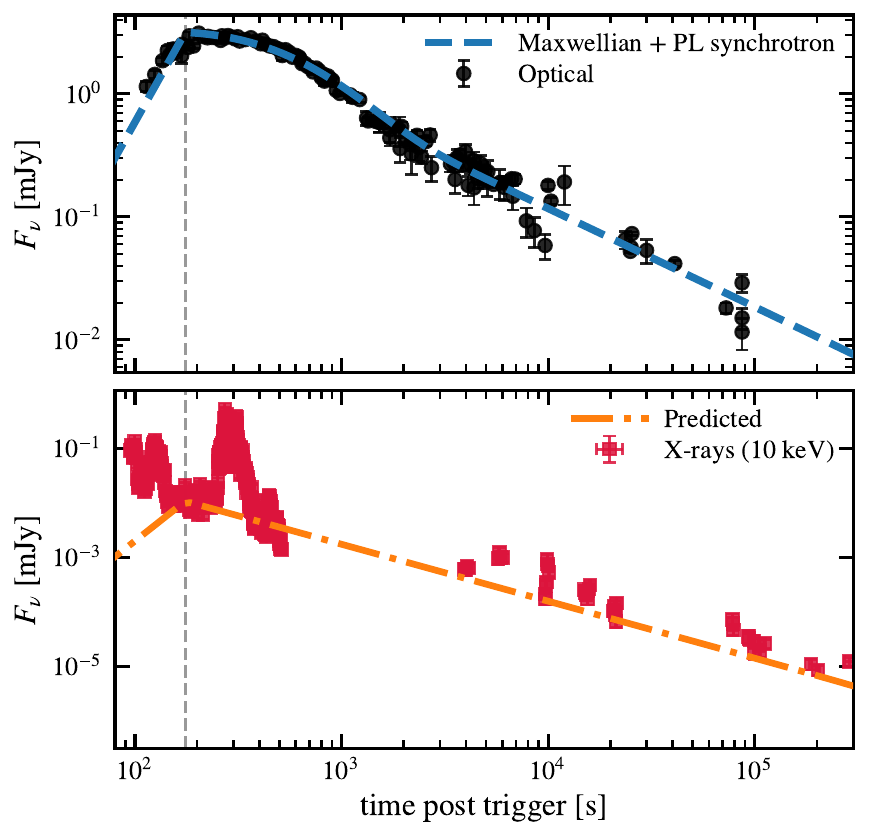}
    \caption{Optical afterglow fit with a hybrid Maxwellian--power-law electron distribution model (top) and the corresponding X-ray prediction based on the optical fit (bottom). The steep decay arises as the synchrotron frequency of thermal electrons sweep through the optical band.}
    \label{fig:MaxPL}
\end{figure}

\struct{Best-fit parameters}
Fitting this model to the optical data at $t > 100$\,s (top panel of Fig.~\ref{fig:MaxPL}; Appendix~\ref{app:hybrid}) yields: non-thermal fraction $\delta = 0.84 \pm 0.02$, electron index $p = 2.05 \pm 0.01$, deceleration time $t_{\rm dec} = 175 \pm 1$\,s, and synchrotron frequency of thermal electrons at the time of deceleration 
\sout{of} 
$\log_{10}(\nu_{\rm th}^0/{\rm Hz}) = 14.43 \pm 0.01$. Provided these constraints, we infer the initial bulk Lorentz factor of the jet $\Gamma_0 \simeq 160$. Thus, the characteristic thermal Lorentz factor of electrons is $\gamma_{\rm th} \approx 900$, assuming equipartition parameter for non-thermal electrons of $\epsilon_e = 0.05$ (see Appendix \ref{app:hybrid}). The comoving magnetic field strength at deceleration is $B' \sim 1.4$\,G, corresponding to the magnetic equipartition parameter of $\epsilon_B \approx 5 \times 10^{-4}$, consistent with constraints from TeV afterglow modeling of several GRBs (see \citet{MN2022} for the review). Given the inferred $\epsilon_B$, we constrain the cooling synchrotron frequency at the time of deceleration to $\nu_c^0 \simeq 4 \times 10^{17}$\,Hz. It allows us to predict the X-ray afterglow, consistent with the \sout{late-time} XRT data (bottom panel of Fig.~\ref{fig:MaxPL}).

\struct{Derived parameters: gamma\_th, B', epsilon\_B}

\struct{X-ray prediction consistent with data}

\section{Discussion}
\label{sec:discussion}

\subsection{From prompt emission to X-ray flares}

\struct{Hard spectrum from optical to MeV -- emission mechanism}
The spectral consistency between the observed optical emission and the extrapolated 8\,keV--40\,MeV spectrum during flare~A implies a power-law extending across five decades in energy.
During the brightest interval, SP1 (28.5–38.5 s), the low-energy photon index is $\alpha_X \sim -1.1$. Within optically thin synchrotron self-Compton models, such a hard spectrum requires dominant inverse-Compton cooling in the Klein–Nishina regime
\citep{Derishev2001,Daigne2011}, implying low magnetic fields ($B^{\prime}\sim$  1–10 G) and electron energies of the TeV scale
\citep{Beniamini2013,Ravasio2019}.
In the subsequent interval SP2 (39.7–49.7 s), the spectrum
softens to $\alpha_X \sim -1.5$, consistent with fast-cooling electrons
and indicating rapid evolution of the emission conditions. Such shallow low-energy slopes remain rare in time-resolved GRB spectra (e.g., \citet{Kaneko2006}).

\struct{X-ray flares = continued internal activity}
The X-ray flares detected during 100--500\,s show spectral shapes similar to MeV prompt emission, namely smoothly broken power laws with $\alpha \approx -0.8$ and $\beta \approx -2.2$ to $-2.4$, but with peak energies shifted to 0.1--4\,keV (Section~\ref{sec:sed}). This supports a physical link between prompt pulses and early X-ray flares as manifestations of continued internal dissipation \citep{Chincarini2010,Margutti2010,Bernardini2011}, distinct from the external-shock-driven optical afterglow.

\subsection{Physics of thermal electrons}
\label{sec:thermal_physics}

\struct{Key clarification: gamma\_th >> Gamma (not intuitive)}
A key aspect of our interpretation requires clarification: the characteristic Lorentz factor of the thermal electrons, $\gamma_{\rm th} \approx 900$, is few times larger than the bulk Lorentz factor of the shock, $\Gamma \approx 160$. This may seem counterintuitive, as one might naively expect thermal electrons to simply have the bulk kinetic energy converted to random motion, yielding $\gamma_{\rm th} \sim \Gamma$.

\struct{PIC simulations: energy equipartition between p and e}
However, particle-in-cell (PIC) simulations of ultra-relativistic collisionless shocks \citep{S08,Sironi2013} demonstrate that energy is approximately equipartitioned between protons and electrons in the downstream region. Electrons are energized far beyond the simple conversion of bulk to thermal energy. The characteristic thermal Lorentz factor for $\delta\approx0.8$ scales as 
\begin{equation}
\gamma_{\rm th} \sim 10^{3} \left(\frac{\epsilon_e}{0.05}\right) \left(\frac{\Gamma}{160}\right),
\end{equation}
where $\epsilon_e$ is the fraction of shock energy deposited into electrons. This energization is a fundamental prediction of collisionless shock physics and has been confirmed in numerous PIC studies 
\citep{Sironi2011,warren}.

\struct{Why this matters: thermal peak in optical, not radio}

\struct{Non-thermal fraction consistent with PIC}
The inferred fraction of energy in non-thermal electrons $\delta \approx 0.8$ implies that the shock acceleration operates, converting most of the electron energy into a power-law tail, while approximately 20\% remains in the thermal pool. This is consistent with PIC simulation results showing that acceleration is efficient but not complete \citep{Sironi2013,warren,Warren_dainotti}.

\section{Conclusions}

\struct{Summary of observations}
We have presented multi-wavelength observations of GRB~250702F, for which the Ond\v{r}ejov D50 robotic telescope obtained high-cadence optical coverage starting just 27.8\,s after trigger -- contemporaneous with the brightest MeV prompt emission pulses.

\struct{Flare A = prompt}
The early optical flare (30--100\,s) is spectrally consistent with the MeV prompt emission, with the extrapolated gamma-ray spectrum correctly predicting the optical flux. This confirms a common origin in internal jet dissipation and provides rare spectral coverage spanning five decades in energy.

\struct{Flare B morphology and interpretation}
The subsequent optical evolution (100--1400\,s) exhibits an unusual rise--plateau--steep-decay morphology that transitions sharply to a standard power-law afterglow at $t \sim 1400$\,s. Standard forward-shock and reverse-shock models fail to explain this morphology (Section~\ref{sec:lcmodel}).

\struct{Thermal electron interpretation}
The steep decay is naturally explained by a hybrid electron distribution at the forward shock, consisting of a thermal (Maxwellian) component smoothly connected to a non-thermal power-law tail. As the shock decelerates, the synchrotron peak of the thermal electrons sweeps through the optical band, producing the observed steep decay. The standard non-thermal afterglow then emerges.

\struct{Parameters and significance}
Model fitting yields a non-thermal energy fraction $\delta = 0.84 \pm 0.02$, electron index $p = 2.05 \pm 0.01$, and characteristic thermal Lorentz factor $\gamma_{\rm th} \approx 900$. The inferred $\epsilon_B \approx 5 \times 10^{-4}$ is consistent with recent TeV afterglow constraints. These observations provide an evidence for thermal electron signatures predicted by PIC simulations, enabled by the combination of early coverage, high cadence, and negligible host extinction.

\begin{acknowledgements}
We thank Emanuele Sobacchi and Pasquale Blasi for the fruitful discussions. BB acknowledges financial support from the Italian Ministry of University and Research (MUR) for the PRIN grant METE under contract no. 2020KB33TP. This research has made use of data obtained through the High Energy Astrophysics Science Archive Research Center Online Service, provided by the NASA/Goddard Space Flight Center, and specifically this work made use of public \textit{Fermi}-GBM and \textit{Fermi}-LAT data. This work made use of data supplied by the UK Swift Science Data Centre at the University of Leicester. The critical early-time observations were obtained with the 0.5-m robotic
telescope D50 at the Astronomical Institute of the
Czech Academy of Sciences in Ondřejov, supported by the project RVO:67985815.
\end{acknowledgements}

\bibliographystyle{aa}
\bibliography{bibliography}

\appendix

\section{Optical observations}
\label{app:photometry}

The Ond\v{r}ejov D50 robotic telescope \citep{d50telescope}, operating under autonomous control via the RTS2 system, began unfiltered observations 27.8\,s after the \textit{Swift} trigger. 
Data were processed in real-time through the automated photometric pipeline \citep{pipeline}. 
Observations switched to photometric filters ($g$, $r$, $i$) approximately 15 minutes after trigger.
Our best optical position for the afterglow, derived from D50 astrometry using Gaia~DR3 reference stars corrected for proper motion and parallax, is
R.A.~$= 14^{\rm h}\,11^{\rm m}\,44\fs64$, Dec.~$= +16\degr\,44\arcmin\,54\farcs48$ (J2000), with uncertainties of $0\farcs035$ in each coordinate.

The optical photometry presented in Table~\ref{tab:photometry} was obtained with the Ond\v{r}ejov D50 telescope and reduced using the \texttt{pyrt} photometric pipeline \citep{pipeline}. This software performs ensemble photometry by fitting precise photometric zeropoints using large numbers of field stars from the Atlas-Refcat2 catalogue \citep{Tonry2018}, simultaneously solving for synthetic flat-field corrections and multi-filter color terms.

Unfiltered (N) observations were calibrated against Atlas-Refcat2, yielding AB magnitudes equivalent to the $r$-band. For the afterglow spectral slope $\beta = 0.52$ (corresponding to $g-r \approx 0.17$\,mag and $r-i \approx 0.13$\,mag including Galactic extinction), the theoretical color correction between unfiltered and $r$-band is $\lesssim 0.01$\,mag. This is consistent with the fitted zeropoint offset of $0.08 \pm 0.05$\,mag (Section~\ref{app:lcfits}), confirming that unfiltered photometry can be treated as $r$-band equivalent. Filtered observations ($g$, $r$, $i$) were calibrated using the respective filter passbands. The late-time GTC upper limit was obtained from a 400\,s $r$-band exposure at $\sim$57 days post-burst.

Extensive optical follow-up was also performed by other facilities; their measurements reported in GCN Circulars \citep[e.g.,][]{GCN40896,GCN40899,GCN40900,GCN40907,GCN40913,GCN40916,GCN41011} are shown in Fig.~\ref{fig:lc} for comparison but not included in the model fitting due to uncertain cross-instrument zeropoints.

\section{High-energy observations and data analysis}
\label{app:highenergy}

GRB~250702F was detected by \textit{Fermi}/GBM and independently triggered by \textit{Swift}/BAT on 2025 July~2 at 21:06:43~UT \citep{GCN40892,GCN40894}. The prompt emission showed a complex temporal structure extending from $\sim$3\,s before to $\sim$80\,s after the BAT trigger, with $T_{90}$ (15--350\,keV) of $63.7 \pm 12.5$\,s.

\textit{Swift} initiated rapid follow-up with XRT and UVOT, with the first XRT observation at 88.5\,s post-trigger. Spectroscopy with GTC/OSIRIS+ measured a redshift of $z = 1.520$ \citep{GCN40901}. The isotropic-equivalent energy $E_{\rm iso} = (9.9 \pm 1.5) \times 10^{52}$\,erg was derived from Konus-\textit{Wind} data \citep{GCN40948}.

\subsection{Fermi GBM}
\label{app:gbm}

We retrieved the \textit{Fermi}/GBM data (8\,keV--40\,MeV) of GRB~250702F from the Fermi GBM Burst Catalog and performed standard data reduction using the Fermi Science Tool \texttt{GTBURST}. We analyzed data from the two sodium iodide (NaI; 8--900\,keV) detectors and one bismuth germanate (BGO; 0.3--40\,MeV) detector with the most favorable observing conditions, namely NaI-3, NaI-6, and BGO-0. The background was modeled by selecting custom time intervals free of source emission, spanning $-110$\,s to $-20$\,s before the trigger and 70\,s to 150\,s after the trigger, which allowed for a stable polynomial background fit.

We fitted the GBM time-resolved spectra during the prompt emission phase (SP1--SP2) in the 10--40000\,keV energy range using a cutoff power-law model. The best-fit parameters are reported in Table~\ref{tab:GBM}. The spectral peak energy, $E_p$, and the cutoff energy, $E_\mathrm{cut}$, are related through $E_p=(2+\alpha_X)E_\mathrm{cut}$. The resulting spectral models are shown in Fig.~\ref{fig:spectra}, where they are extrapolated to the optical energy range for comparison with the D50 observations.

\begin{table}[ht]
\renewcommand{\arraystretch}{1.4}
\centering
\caption{Spectral parameters for cutoff power-law model fits to prompt emission spectra.}
\label{tab:GBM}
\begin{tabular}{cccccc}
\hline
SP & $t_{\mathrm{start}}$ & $t_{\mathrm{end}}$ &
$F_\mathrm{10-40000\,keV}$ & $\alpha_X$ & $E_{\mathrm{cut}}$ \\
 & [s] & [s] & [erg\,cm$^{-2}$\,s$^{-1}$] & & [keV] \\
\hline
1 & 28.5 & 38.5
  & $(3.4^{+0.5}_{-0.4}) \times 10^{-7}$
  & $-1.14_{-0.06}^{+0.07}$
  & $885_{-222}^{+335}$ \\
2 & 39.7 & 49.7
  & $(2.6^{+1.0}_{-0.7}) \times 10^{-7}$
  & $-1.49_{-0.07}^{+0.07}$
  & $> 2000$ \\
\hline
\end{tabular}
\end{table}

\subsection{Swift XRT and BAT}
\label{app:xrt}

We retrieved the XRT and BAT light curves from the burst analyser web tool provided by the \textit{Swift} Science Data Center \citep{2010A&A...519A.102E}. XRT spectral files in both WT and PC modes, along with corresponding background files, redistribution matrices, and ancillary response files, were obtained using the automated online spectral analysis tool \citep{2009MNRAS.397.1177E}.

BAT spectra were extracted using the HEASOFT software package (v6.33.1). The BAT event files were retrieved from the \textit{Swift} data archive and processed with the \texttt{batgrbproduct} pipeline. Spectral files were generated using \texttt{batbinevt} and corrected for systematic uncertainties with \texttt{batupdatephakw} and \texttt{batphasyserr}. Response matrices were produced using \texttt{batdrmgen}.

We performed time-resolved joint XRT and BAT spectral fits for each time bin (SP3--SP8) in the 0.3--150\,keV band using a smoothly broken power-law (sBPL) model. Since the optical emission shows no rebrightening during the X-ray flares, we required the extrapolated spectrum to lie below the observed optical flux by imposing a prior on the flux in the 1.8--3.1\,eV range. The results are presented in Table~\ref{tab:xray}.

\begin{table*}[ht]
\centering
\renewcommand{\arraystretch}{1.4}
\caption{Spectral parameters for sBPL model fits to X-ray flare spectra.}
\label{tab:xray}
\begin{tabular}{ccccccc}
\hline
SP & $t_{\mathrm{start}}$ [s] & $t_{\mathrm{end}}$ [s] &
$F_\mathrm{opt}$ [erg\,cm$^{-2}$\,s$^{-1}$] &
$E_p$ [keV] & $\alpha_X$ & $\beta_X$ \\
\hline
3 & 97.1 & 118.1 & $>1.5 \times 10^{-13}$ & $1.7^{+0.3}_{-0.2}$ & $-0.7^{+0.3}_{-0.2}$ & $-2.2^{+0.1}_{-0.1}$ \\
4 & 119.6 & 151.6 & $>2.8 \times 10^{-13}$ & $2.4^{+1.5}_{-0.6}$ & $-0.8^{+0.4}_{-0.2}$ & $-2.08^{+0.04}_{-0.09}$ \\
5 & 153.4 & 242.4 & $>1.9 \times 10^{-13}$ & $0.4^{+0.1}_{-0.1}$ & $>-0.8$ & $-2.26^{+0.04}_{-0.05}$ \\
6 & 243.4 & 298.4 & $>3.8 \times 10^{-12}$ & $4.3^{+1.3}_{-0.6}$ & $-0.8^{+0.3}_{-0.2}$ & $-2.20^{+0.05}_{-0.07}$ \\
7 & 299.8 & 394.8 & $>6.7 \times 10^{-13}$ & $0.9^{+0.1}_{-0.1}$ & $-0.8^{+0.3}_{-0.2}$ & $-2.20^{+0.05}_{-0.07}$ \\
8 & 396.2 & 501.2 & $>3.6 \times 10^{-13}$ & $0.16^{+0.07}_{-0.06}$ & $>-0.5$ & $-2.46^{+0.04}_{-0.04}$ \\
\hline
\end{tabular}
\end{table*}

\subsection{Fermi LAT}
\label{app:lat}

We performed unbinned likelihood analysis of \textit{Fermi}/LAT data for GRB~250702F ($t_0 = 773183208.102$\,s MET) extending to $t_0 + 1$\,ks, in the energy range 0.1--10\,GeV, using the \texttt{GTBURST} software. We selected a region of interest covering $12^\circ$ around the source location (R.A.$= 212.93^\circ$, Dec.$= 16.69^\circ$, J2000). A standard zenith angle cut of $100^\circ$ was applied to remove Earth-limb contamination. We used the \texttt{P8R3\_TRANSIENT020} event class with corresponding instrument response functions, and included the isotropic particle background and 4FGL catalog sources with fixed normalization.

The source is not significantly detected in any time bin (TS $< 25$). A marginal excess (TS $\sim 18$) is observed in SP8, with the highest-energy photon ($\sim$4.5\,GeV) having association probability $> 0.9$. The 95\% confidence upper limits are reported in Table~\ref{tab:lat}.

\begin{table}[ht]
\renewcommand{\arraystretch}{1.4}
\centering
\caption{LAT time-resolved analysis results.}
\label{tab:lat}
\begin{tabular}{ccc}
\hline
SP (time interval) & Energy & Flux upper limit (TS) \\
 & [GeV] & [erg\,cm$^{-2}$\,s$^{-1}$] \\
\hline
1 (28.5--38.5\,s) & 0.1--1.0 & No photons \\
2 (39.7--49.7\,s) & 0.1--1.0 & $<2.6\times10^{-8}$ (0) \\
3 (97.1--118.1\,s) & 0.1--1.0 & No photons \\
4 (119.6--151.6\,s) & 0.1--1.0 & $<1.0\times10^{-8}$ (3) \\
5 (153.4--242.4\,s) & 0.1--1.0 & $<2.7\times10^{-9}$ (0) \\
6 (243.4--298.4\,s) & 0.1--1.0 & $<6.2\times10^{-9}$ (1) \\
7 (298.4--394.8\,s) & 0.1--1.0 & $<2.9\times10^{-9}$ (0) \\
8 (396.2--501.2\,s) & 0.1--1.0 & $<4.1\times10^{-9}$ (18) \\
8 (396.2--501.2\,s) & 0.1--5.0 & $<7.0\times10^{-9}$ (18) \\
\hline
\end{tabular}
\end{table}

\section{Empirical optical light curve modeling}
\label{app:lcfits}

We fitted the optical light curve at $t > 100$\,s (Fig.~\ref{fig:lc}) with a model consisting of two components joined at a sharp break time $t_b$. For $t < t_b$, we use a double hyperbola that smoothly transitions through three power-law segments with asymptotic slopes $\alpha_1$ (rise), $\alpha_2$ (plateau), and $\alpha_3$ (steep decay), with vertex times $t_{v,1}$ and $t_{v,2}$ controlling the transitions. For $t > t_b$, the light curve follows a standard forward-shock power-law decay constrained by closure relations: the temporal slope $\alpha_4 = 3(p-1)/4$ and spectral slope $\beta = (p-1)/2$ are both determined by the electron index $p$, appropriate for $\nu < \nu_c$ in a homogeneous medium.

We adopted uniform priors on the model parameters (see Table~\ref{tab:lcfit}) and sampled the posterior distribution using MCMC with the \texttt{emcee} package \citep{2013PASP..125..306F}, fitting 80 D50 photometric points. A zeropoint offset between unfiltered and $r$-band observations was included as a nuisance parameter; the fitted value ($0.08 \pm 0.05$\,mag) is consistent with zero. The fit achieves $\chi^2/\mathrm{dof} = 82/68 = 1.2$.

\begin{table}[ht]
\renewcommand{\arraystretch}{1.4}
\centering
\caption{Model parameters and prior distributions for the empirical light curve fit. The last column reports the median and 68\% credible intervals of the posterior distribution.}
\label{tab:lcfit}
\begin{tabular}{l c c}
\hline
Parameter & Prior & Posterior \\
\hline
$\alpha_1$ (rise) & $\mathcal{U}(-4,\,0)$ & $-3.1^{+0.9}_{-0.6}$ \\
$\alpha_2$ (plateau) & $\mathcal{U}(-1,\,1.5)$ & $-0.17^{+0.18}_{-0.29}$ \\
$\alpha_3$ (steep decay) & $\mathcal{U}(0.9,\,3)$ & $1.8^{+0.8}_{-0.6}$ \\
$\log_{10}(t_{v,1}/\mathrm{s})$ & $\mathcal{U}(1.5,\,2.5)$ & $2.06^{+0.05}_{-0.04}$ \\
$\log_{10}(t_{v,2}/\mathrm{s})$ & $\mathcal{U}(2.0,\,3.5)$ & $2.94^{+0.19}_{-0.18}$ \\
$\log_{10}(t_b/\mathrm{s})$ & $\mathcal{U}(3.08,\,3.23)$ & $3.15^{+0.02}_{-0.02}$ \\
$p$ & $\mathcal{U}(1.8,\,2.5)$ & $2.05 \pm 0.04$ \\
\hline
\multicolumn{3}{c}{Derived parameters} \\
\hline
$\alpha_4 = 3(p-1)/4$ & -- & $0.79 \pm 0.03$ \\
$\beta = (p-1)/2$ & -- & $0.52 \pm 0.02$ \\
\hline
\end{tabular}
\end{table}

\section{Two-component fit of the optical light curve}
\label{app:twocomp}

\begin{figure}[ht]
    \centering
    \includegraphics[width=\columnwidth]{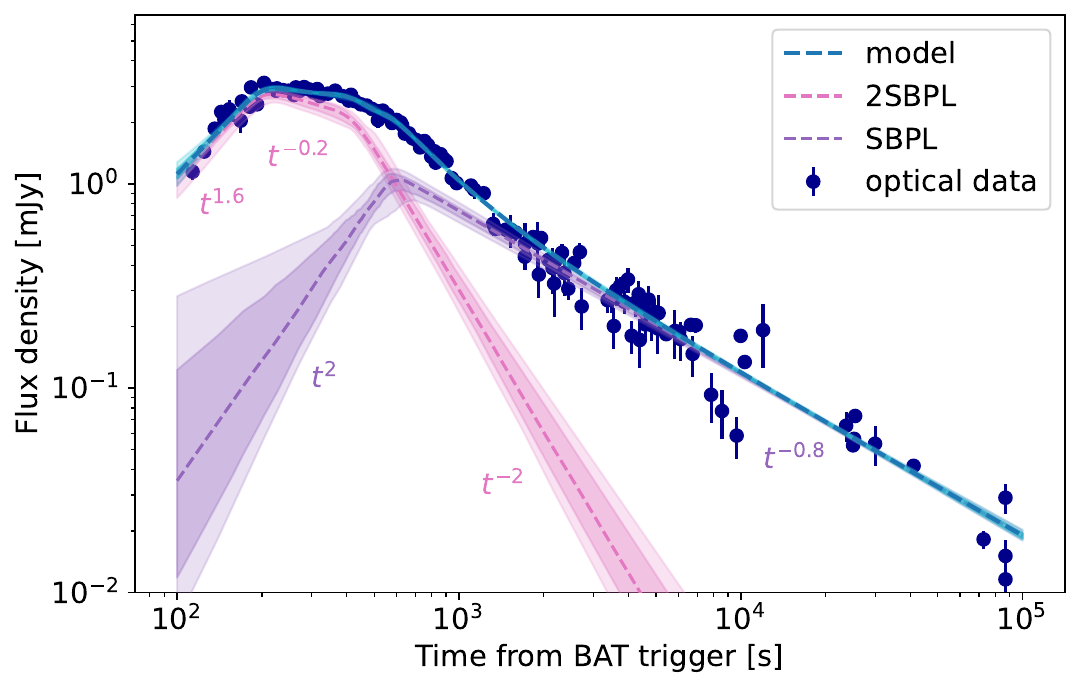}
    \caption{Two-component (RS + FS) fit to the optical light curve. The inferred RS peak time does not coincide with the FS peak, contrary to the expectation that both components originate from the same deceleration radius.}
    \label{fig:twocomp}
\end{figure}

To test the reverse-shock interpretation, we fitted the optical light curve (>100 s) with a two-component model: a smoothly-broken power law (SBPL), describing the underlying forward-shock (FS), and a superposed double smoothly-broken power law (2SBPL), representing the reverse-shock (RS). Both components were modeled with independent break times, normalizations and temporal indices, as shown in Fig. \ref{fig:twocomp}. We adopted uniform priors on the model parameters (see Table~\ref{tab:two-component}), and sampled the posterior distribution using a Markov Chain Monte Carlo method implemented with the \texttt{emcee} package \citep{2013PASP..125..306F}. The fit results are reported in Table~\ref{tab:two-component}.

The key prediction of RS models is that the RS and FS should peak at approximately the same time, since both arise from the deceleration of the ejecta at the same radius. In our fits, however, the inferred RS peak time ($t_{\rm peak,RS} \sim 400$--$600$\,s) is significantly later than the FS peak ($t_{\rm peak,FS} \sim 100$--$200$\,s, corresponding to the deceleration time). This temporal offset cannot be reconciled with standard RS theory.

Furthermore, the extended plateau phase ($\alpha \approx 0$ between 200--600\,s) requires fine-tuning in the two-component model: the RS decay and FS rise must nearly cancel over an extended period, which is not a natural outcome of the physics. In contrast, the thermal electron model (Appendix~\ref{app:hybrid}) produces the plateau naturally as the thermal synchrotron peak frequency passes through the optical band.

\begin{table}[ht]
\centering
\renewcommand{\arraystretch}{1.4}
\caption{Model parameters and prior distributions adopted in the empirical fit of the optical light curve. 
For the 2SBPL component, an ordering prior $\log_{10}(t_{b,1}) < \log_{10}(t_{b,2})$ was imposed. The last column reports the median and the 68\% credible intervals of the posterior distribution.}
\label{tab:two-component}
\begin{tabular}{c c c c}
\hline
Model & Parameter & Prior & Posterior\\
\hline
\multirow{6}{*}{2SBPL}
 & $\log_{10}(A_1/\mathrm{mJy})$   & $\mathcal{U}(-5,\,3)$ & $0.31^{+0.03}_{-0.05}$\\
 & $\alpha_1$        & $\mathcal{U}(-5,\,5)$ & $-1.64^{+0.13}_{-0.14}$\\
 & $\alpha_2$        & $\mathcal{U}(-5,\,5)$ & $0.19^{+0.05}_{-0.04}$\\
 & $\alpha_3$        & $\mathcal{U}(-5,\,5)$ & $2.06^{+0.18}_{-0.16}$\\
 & $\log_{10}(t_{b,1}/\mathrm{s})$ & $\mathcal{U}(2.0,\,3.3)$ & $2.30^{+0.02}_{-0.02}$\\
 & $\log_{10}(t_{b,2}/\mathrm{s})$ & $\mathcal{U}(2.0,\,3.3)$ & $2.62^{+0.03}_{-0.02}$\\
\hline
\multirow{4}{*}{SBPL}
 & $\log_{10}(A_2/\mathrm{mJy})$   & $\mathcal{U}(-5,\,3)$  & $0.01^{+0.03}_{-0.04}$\\
 & $\alpha_4$        & $\mathcal{U}(-5,\,0)$  & $-1.98^{+0.71}_{-0.59}$\\
 & $\alpha_5$        & $\mathcal{U}(0,\,5)$   & $0.79^{+0.02}_{-0.02}$\\
 & $\log_{10}(t_{b,3}/\mathrm{s})$ & $\mathcal{U}(2.7,\,4.0)$ & $2.77^{+0.03}_{-0.02}$\\
\hline
\end{tabular}
\end{table}

\section{Hybrid electron distribution model}
\label{app:hybrid}

To model the optical light curve at $t \gtrsim 100$\,s, we adopt a hybrid electron energy distribution consisting of a thermal (Maxwellian) component and a non-thermal power-law tail, following \citet[][hereafter GS09]{GS2009}. Electrons crossing the relativistic forward shock populate a relativistic Maxwellian at low energies, with a fraction subsequently accelerated into a power-law tail. The electron distribution is:
\begin{equation}
N_e(\gamma,\gamma_{\rm th}) =
\begin{cases}
C\,N_e^{\rm th}(\gamma,\gamma_{\rm th}), & \gamma \le \gamma_m, \\
C\,N_e^{\rm th}(\gamma_m,\gamma_{\rm th})
\left(\dfrac{\gamma}{\gamma_m}\right)^{-p},
& \gamma > \gamma_m,
\end{cases}
\end{equation}
where
\begin{equation}
N_e^{\rm th}(\gamma,\gamma_{\rm th})
= \dfrac{\gamma^2}{2\gamma_{\rm th}^3}
\exp\!\left(-\dfrac{\gamma}{\gamma_{\rm th}}\right)
\end{equation}
is the relativistic Maxwellian distribution. Here $\gamma_{\rm th}$ is the characteristic thermal Lorentz factor, $\gamma_m$ is the minimum Lorentz factor of the power-law tail, $p$ is the power-law index, and $C$ is a normalization constant.

The distribution is parameterized by: (i) the fraction $\delta$ of electron energy in the non-thermal tail, (ii) the characteristic thermal Lorentz factor $\gamma_{\rm th}$, and (iii) characteristic Lorentz factor of electrons $\gamma_c$ above which electrons loose their energy in a dynamical time-scale. We assume slow cooling regime ($\gamma_c > \gamma_m$) throughout, where $\gamma_m$ is the minimum Lorentz factor of non-thermal electrons.

Our phenomenological model parameters are: (1) $\delta$, (2) the thermal synchrotron frequency at deceleration $\nu_{\rm th}^0$, (3) the cooling frequency $\nu_c^0$, at the deceleration time (4) the power-law index $p$ of non-thermal electrons, and (5) the deceleration time $t_{\rm dec}$. For $t < t_{\rm dec}$, we assume $F_\nu \propto t^3$ (coasting phase). At later times, $\nu_{\rm th}(t) = \nu_{\rm th}^0 (t/t_{\rm dec})^{-3/2}$.

Best-fit parameters are: $\delta = 0.84 \pm 0.02$, $p = 2.05 \pm 0.01$, $t_{\rm dec} = 175 \pm 1$\,s, and $\log_{10}(\nu_{\rm th}^0/\mathrm{Hz}) = 14.43 \pm 0.01$.

The initial bulk Lorentz factor is (homogeneous medium):
\begin{equation}
\Gamma_0 \simeq 160\,E_{\rm iso,53}^{1/8}\,n_0^{-1/8}\,\eta_{-1}^{-1/8}\,t_{\rm dec,175}^{-3/8}(1+z)^{3/8}.
\end{equation}
Assuming $\epsilon_e = 0.05$ and $\Gamma \simeq 160$, we infer $\gamma_{\rm th} \simeq 900$.

The comoving magnetic field strength is $B' \simeq 1.4$\,G, corresponding to:
\begin{equation}
\epsilon_B \simeq 5 \times 10^{-4}
\left(\frac{B'}{1.4\,\mathrm{G}}\right)^2
\left(\frac{\Gamma}{160}\right)^{-2}
\left(\frac{n}{1\,\mathrm{cm^{-3}}}\right)^{-1}.
\end{equation}
The cooling frequency at deceleration is $\nu_c^0 \simeq 4 \times 10^{17}$\,Hz, evolving as $\nu_c(t) = \nu_c^0 (t/t_{\rm dec})^{-1/2}$ (homogeneous medium).

\section{Archival GRB comparison}
\label{app:archival}

We searched for similar rise--plateau--steep-decay structures in archival GRB optical light curves. GRB~161023A \citep{2018A&A...620A.119D} exhibits morphologically similar early behavior. However, detailed analysis reveals consistency with a steeper electron index $p \approx 2.8$, which explains both the temporal decay slopes and the optical-to-X-ray SED through standard forward-shock closure relations without invoking thermal electrons. A similar structure appears at lower signal-to-noise in GRB~201015A (FRAM, unpublished), but data quality is insufficient for conclusive spectral constraints.

The key distinction of GRB~250702F is its negligible host extinction ($A_V \approx 0$), which directly constrains the intrinsic spectral slope and firmly locks $p = 2.05$. This removes the $\beta$--extinction degeneracy that can otherwise mask the true electron index. The combination of well-sampled early optical coverage and clean SED constraints makes GRB~250702F well suited for testing the thermal electron scenario.

\begin{table*}
\caption{Photometric observations of GRB 250702F.}\label{tab:photometry}
\centering
\small
\begin{tabular}{cccccc|cccccc}
\hline\hline
$t_{\rm mid}$ & $t_{\rm exp}$ & Magnitude & Error & Filter & Telescope &
$t_{\rm mid}$ & $t_{\rm exp}$ & Magnitude & Error & Filter & Telescope \\
(s) & (s) & (AB mag) & (AB mag) & & &
(s) & (s) & (AB mag) & (AB mag) & & \\
\hline
33.5 & 10 & 15.423 & 0.050 & N & D50 & 727.7 & 20 & 15.918 & 0.047 & N & D50 \\
44.7 & 10 & 13.860 & 0.013 & N & D50 & 754.3 & 20 & 15.833 & 0.048 & N & D50 \\
57.1 & 10 & 14.833 & 0.028 & N & D50 & 775.7 & 20 & 15.984 & 0.055 & N & D50 \\
68.3 & 10 & 15.528 & 0.052 & N & D50 & 796.9 & 20 & 16.065 & 0.056 & N & D50 \\
79.6 & 10 & 16.290 & 0.097 & N & D50 & 818.1 & 20 & 15.984 & 0.052 & N & D50 \\
90.8 & 10 & 16.680 & 0.142 & N & D50 & 839.4 & 20 & 16.122 & 0.059 & N & D50 \\
102.1 & 10 & 16.985 & 0.186 & N & D50 & 860.7 & 20 & 16.075 & 0.056 & N & D50 \\
113.4 & 10 & 16.350 & 0.104 & N & D50 & 881.9 & 20 & 16.160 & 0.062 & N & D50 \\
124.6 & 10 & 15.942 & 0.074 & N & D50 & 903.2 & 20 & 16.184 & 0.061 & N & D50 \\
135.9 & 10 & 15.746 & 0.063 & N & D50 & 981.6 & 120 & 16.411 & 0.045 & $g$ & D50 \\
147.1 & 10 & 15.646 & 0.055 & N & D50 & 1103 & 120 & 16.433 & 0.038 & $r$ & D50 \\
158.3 & 10 & 15.502 & 0.049 & N & D50 & 1224 & 120 & 16.506 & 0.053 & $i$ & D50 \\
169.7 & 10 & 15.432 & 0.046 & N & D50 & 1347 & 120 & 16.962 & 0.067 & $g$ & D50 \\
180.9 & 10 & 15.455 & 0.048 & N & D50 & 1468 & 120 & 16.970 & 0.074 & $r$ & D50 \\
192.1 & 10 & 15.412 & 0.047 & N & D50 & 1589 & 120 & 16.985 & 0.084 & $i$ & D50 \\
203.4 & 10 & 15.167 & 0.037 & N & D50 & 1712 & 120 & 17.301 & 0.083 & $g$ & D50 \\
214.7 & 10 & 15.246 & 0.041 & N & D50 & 1833 & 120 & 17.067 & 0.064 & $r$ & D50 \\
225.9 & 10 & 15.278 & 0.041 & N & D50 & 1955 & 120 & 17.066 & 0.080 & $i$ & D50 \\
237.2 & 10 & 15.221 & 0.039 & N & D50 & 2077 & 120 & 17.344 & 0.081 & $g$ & D50 \\
248.5 & 10 & 15.152 & 0.037 & N & D50 & 2199 & 120 & 17.359 & 0.080 & $r$ & D50 \\
259.7 & 10 & 15.283 & 0.040 & N & D50 & 2320 & 120 & 17.237 & 0.092 & $i$ & D50 \\
270.9 & 10 & 15.254 & 0.039 & N & D50 & 2443 & 120 & 17.705 & 0.111 & $g$ & D50 \\
282.2 & 10 & 15.239 & 0.039 & N & D50 & 2564 & 120 & 17.384 & 0.073 & $r$ & D50 \\
293.5 & 10 & 15.243 & 0.040 & N & D50 & 2685 & 120 & 17.235 & 0.101 & $i$ & D50 \\
304.7 & 10 & 15.231 & 0.038 & N & D50 & 3369 & 120 & 17.849 & 0.127 & $g$ & D50 \\
321.2 & 20 & 15.292 & 0.030 & N & D50 & 3490 & 120 & 17.811 & 0.137 & $r$ & D50 \\
342.5 & 20 & 15.276 & 0.029 & N & D50 & 3612 & 120 & 17.689 & 0.152 & $i$ & D50 \\
363.7 & 20 & 15.241 & 0.028 & N & D50 & 3735 & 120 & 17.652 & 0.097 & $g$ & D50 \\
385.0 & 20 & 15.333 & 0.030 & N & D50 & 3856 & 120 & 17.856 & 0.116 & $r$ & D50 \\
406.2 & 20 & 15.372 & 0.031 & N & D50 & 3977 & 120 & 17.566 & 0.127 & $i$ & D50 \\
427.5 & 20 & 15.367 & 0.031 & N & D50 & 4100 & 120 & 18.274 & 0.167 & $g$ & D50 \\
448.7 & 20 & 15.408 & 0.032 & N & D50 & 4221 & 120 & 17.886 & 0.134 & $r$ & D50 \\
470.0 & 20 & 15.426 & 0.032 & N & D50 & 4343 & 120 & 17.746 & 0.153 & $i$ & D50 \\
491.2 & 20 & 15.483 & 0.033 & N & D50 & 4465 & 120 & 17.949 & 0.126 & $g$ & D50 \\
512.5 & 20 & 15.516 & 0.034 & N & D50 & 4587 & 120 & 18.136 & 0.158 & $r$ & D50 \\
536.4 & 20 & 15.493 & 0.035 & N & D50 & 4708 & 120 & 17.803 & 0.154 & $i$ & D50 \\
557.7 & 20 & 15.520 & 0.034 & N & D50 & 4831 & 120 & 18.145 & 0.150 & $g$ & D50 \\
578.9 & 20 & 15.622 & 0.039 & N & D50 & 4952 & 120 & 17.982 & 0.134 & $r$ & D50 \\
600.2 & 20 & 15.585 & 0.037 & N & D50 & 5073 & 120 & 18.164 & 0.228 & $i$ & D50 \\
621.4 & 20 & 15.612 & 0.038 & N & D50 & 5439 & 605 & 18.230 & 0.054 & $g$ & D50 \\
642.7 & 20 & 15.753 & 0.041 & N & D50 & 6045 & 605 & 18.185 & 0.062 & $r$ & D50 \\
664.0 & 20 & 15.795 & 0.044 & N & D50 & 6651 & 605 & 18.145 & 0.069 & $i$ & D50 \\
685.2 & 20 & 15.832 & 0.044 & N & D50 & 5.10e6 & 400 & 24.660 & 0.062 & $r$ & GTC \\
706.5 & 20 & 15.848 & 0.046 & N & D50 & & & & & & \\
\hline
\end{tabular}
\end{table*}

\end{document}